\documentclass[aps,pra,twocolumn,superscriptaddress]{revtex4}
\usepackage{graphicx}% Include figure files
\usepackage{dcolumn}% Align table columns on decimal point
\usepackage{amsmath,bm}% bold math
\usepackage{amssymb,amsmath,amsbsy,amsgen,amsfonts}
\usepackage{dcolumn}
\usepackage{amsthm}
\usepackage{latexsym}
\usepackage{array}
\usepackage{amstext}
\usepackage{epsfig}
\usepackage{color}
\usepackage{units}
\usepackage{calrsfs}

\newcommand{\be}{\begin{equation}}
\newcommand{\ee}{\end{equation}}
\newcommand{\ba}{\begin{array}}
\newcommand{\ea}{\end{array}}
\newcommand{\bqa}{\begin{eqnarray}}
\newcommand{\eqa}{\end{eqnarray}}

\begin{document}

\title{Broadband field enhancement and giant nonlinear effects \\in terminated unidirectional plasmonic waveguides}

\author{S. Ali Hassani Gangaraj} \email{ali.gangaraj@gmail.com}
\address{School of Electrical and Computer Engineering, Cornell University, Ithaca, NY 14853, USA}

\author{Boyuan Jin}\email{byjin328@huskers.unl.edu}
\address{Department of Electrical and Computer Engineering, University of Nebraska-Lincoln, Lincoln, NE 68588, USA}

\author{Christos Argyropoulos} \email{christos.argyropoulos@unl.edu}
\address{Department of Electrical and Computer Engineering, University of Nebraska-Lincoln, Lincoln, NE 68588, USA}

\author{Francesco Monticone} \email{francesco.monticone@cornell.edu}
\address{School of Electrical and Computer Engineering, Cornell University, Ithaca, NY 14853, USA}

\date{\today}

\begin{abstract}

%This paper presents the analysis and application of terminated one-way plasmonic waveguiding structure in enhancing high harmonic nonlinear generation processes. For this purpose, the complex behavior of a terminated transmission line supporting unidirectional nonreciprocal surface plasmon-polaritons is theoretically analyzed and exact closed form modal solutions is provided. The outcome exhibit a remarkable electric field confinement around the termination point. This significant large electric field enhancement can lead to substantial boosting in different nonlinear optical phenomena. Inspired by this approach, a practical geometry is proposed that is composed of a silicon (Si) layer sandwiched in between two oppositely biased indium antimonide (InSb) semiconductors. It is shown that the InSb-Si-InSb interfaces support two coupled unidirectional surface waves that can be completely stopped by an opaque medium termination. By considering the third-order susceptibility in the nonlinear dielectric polarization density function of Si, the nonlinear, nonlocal non-Hermitian and dispersive behavior of the such waveguide is explored. It is shown that the nonlinear response is dramatically enhanced with the proposed terminated nonreciprocal surface waveguiding system. Several potential applications are envisioned based on the proposed concept, such as efficient frequency generators, nonlinear sensors and all-optical wave mixers.

Unidirectional wave propagation in nonreciprocal structures enables exciting opportunities to control and enhance wave-matter interactions in extreme ways. Within this context, here we investigate the possibility of using terminated unidirectional plasmonic waveguides %, suitably coupled to nonlinear elements, 
to enhance typically weak nonlinear effects by orders of magnitude. We theoretically demonstrate remarkable levels of electric field enhancement and confinement (field hot-spots) when the unidirectional waveguiding structure is terminated with a suitable boundary that fully stops the one-way mode. Such a large field enhancement, originating from a nonresonant effect, is fundamentally different from the narrow-band field concentration effects in resonant plasmonic structures. Instead, it is analogous to the broadband response of plasmonic tapers, but without the need for any adiabatic impedance matching. We show that this effect can indeed lead to a substantial boosting of nonlinear light-matter interactions, exemplified by an improvement of several orders of magnitude in the third-harmonic-generation efficiency, which is of large significance for several applications. More broadly, our findings show the potential of extreme nonreciprocal configurations for enhanced wave-matter interactions.

%To leverage this effect, we propose and numerically demonstrate a realistic wave-guiding geometry composed of a silicon (Si) layer sandwiched between biased indium antimonide (InSb) semiconductors. The InSb-Si-InSb interfaces support two coupled unidirectional surface waves, which propagate for a certain length before reaching an opaque, metallic termination that fully stops them, as no wave can propagate along the surface of the termination layer and no backward wave is allowed. We consider the third-order susceptibility in the nonlinear dielectric polarization-density function of Si, and we study the third-harmonic generation provided by this complex configuration involving materials that are, at the same time, nonlinear, nonlocal, dissipative, and dispersive. We theoretically and numerically show that the nonlinear response is indeed dramatically enhanced, with an order-of-magnitude improvement in the third-harmonic-generation conversion efficiency. Several potential applications are envisioned based on the proposed concept, such as efficient frequency generators, nonlinear sensors, and all-optical wave mixers. More generally, we believe our findings show the potential of extreme nonreciprocal configurations for enhanced wave-matter interactions.

\end{abstract}

\maketitle

%%%%%%%%%%%%%%%%%%%%%%%%%%%%%%%%%%%%%%%%%%%%%%
%%%%%%%%%%%%%%%%%%%%%%%%%%%%%%%%%%%%%%%%%%%%%%
%%%%%%%%%%%%%%%%%%%%%%%%%%%%%%%%%%%%%%%%%%%%%%

\section{Introduction}

Nonlinear light-matter interactions are at the basis of a large variety of classical and quantum optical devices, and are used by scientists and engineers to generate new light frequencies, perform laser diagnostics, and advance quantum computing, among many other applications \cite{Boyd}. Nonlinear effects depend on the powers of the local electric field $E$, with the nonlinear polarization density, in time domain, given by: ${P}\left( t \right) = {\varepsilon _0}\left( {\chi \,{E}\left( t \right) + {\chi ^{\left( 2 \right)}}\,{{E}^2}\left( t \right) + {\chi ^{\left( 3 \right)}}\,{{E}^3}\left( t \right) +  \ldots } \right)$, where $\chi$ is the linear susceptibility and $\chi^{(n)}$ the nonlinear susceptibilities. Since $\chi^{(n)} \ll \chi$ in natural materials, %in order to achieve a large nonlinear response, a nonlinear device should either be sufficiently long, so that light interacts with matter for a longer time, or the local field intensity needs to be enhanced to values much larger than the incident field intensity. 
extremely high incident light intensities, enhanced local fields in narrow-band cavities, or long propagation distances in bulky nonlinear crystals are required to produce detectable optical nonlinear effects. The realization of compact broadband nonlinear devices (wavelength-scale or smaller) is indeed a fundamental challenge in modern photonics and nano-optics.

Two general strategies are typically used to locally enhance electromagnetic fields: (i) localized resonances and (ii) slow-light effects accompanied by adiabatic impedance matching. Resonances, for example localized surface plasmon resonances in plasmonic nanostructures and metasurfaces, or more complex Fano resonances and bound states in the continuum, can dramatically increase the local field, producing field hot-spots that may be used to boost linear and nonlinear effects \cite{Non_11,Non_31,C1,C2,Non_21,Non_22,Non_32,Kivshar,Monticone-Embedded,Mario_trapping,Hsu,Minkov}. However, the resonant nature of these platforms makes them very sensitive to dissipative processes and, more importantly, reduces the bandwidth over which the desired effect is obtained. An alternative approach is to use elongated plasmonic tapers \cite{Stockman,Non_15,C3}, which support surface plasmon-polariton (SPP) modes with decreasing wavelength and group velocity as they propagate toward the taper tip. For long adiabatic tapers, the energy carried by the SPP tends to accumulate at the tip, producing an intense field hot-spot. Despite the change in geometry seen by the propagating SPP, resulting in a change in wave impedance, reflections are minimized over a broad bandwidth (broadband impedance matching) due to the adiabatic transition. In this way, a broadband signal can be focused at the tip of a plasmonic taper, producing a ultra-large field hot-spot without the need for a local resonance. However, this effect is possible only if impedance matching is ensured at any distance from the tip, which requires very long adiabatically tapered structures.

Here, we propose a novel strategy to fundamentally break these conventional trade-offs between field enhancement, bandwidth, and size -- with the final goal of realizing boosted nonlinear effects -- based on exploiting the extreme response of nonreciprocal plasmonic platforms supporting inherently unidirectional modes. Nonreciprocal plasmas and plasmonic materials, obtained by breaking Lorentz reciprocity via an external magnetic field (gyrotropic materials), have been the subject of extensive research for decades. A distinctive effect of nonreciprocity in this context is the existence of frequency ranges where surface plasmon-polaritons are unidirectional in the plane orthogonal to the bias (Voigt configuration) \cite{FM_Nature}. In other words, the SPP dispersion diagram is markedly asymmetric and, in certain frequency windows, SPPs are allowed to propagate along a certain direction but not in the opposite direction. Although this effect has been known for several decades \cite{1972,Seshadri,Ishimaru}, only recently it has been shown that unidirectional SPPs on nonreciprocal plasmonic structures can be divided in two classes with distinct topological properties: (i) Topological SPPs, whose unidirectionality is an intrinsic property arising from the topological nature of the bulk modes  \cite{Mario_Chern,Soljacic_1,Shen_1,Z_Yu,sink,soljacic_2,Hassani_1,Zubin,Davoyan,PRL_Exceptional}. These SPPs are supported by interfaces between a biased plasmonic material and an opaque medium (e.g., a metal) and exist within the upper bulk-mode bandgap of the plasma. (ii) Unidirectional surface magneto-plasmons, which are supported by interfaces between a biased plasmonic material and a transparent medium, and exist within the lower bulk-mode bandgap. The unidirectionality of these SPPs is a manifestation of strong nonreciprocity, and not a topological property \cite{Optica,diffractionless,Fan,TAP,Tsakmakidis}. Indeed, it has been recently recognized that these SPPs lose their strict unidirectionality if nonlocal effects are properly included in the material model \cite{Optica,Fan}, while they remain strongly asymmetrical. 

Unidirectional SPPs of both classes can be used to achieve significant field enhancements in configurations where the unidirectional mode is fully stopped at a suitable termination. Since the SPP cannot reflect back, its energy accumulates at the termination, forming an intense field hot-spot, and is eventually dissipated in the form of heat (or radiation loss). Terminated unidirectional wave-guiding structures -- and their counterintuitive electromagnetic response -- were originally studied by Barzilai and Ishimaru, among others, in the 1960s \cite{Ishimaru,Barzilai}, and are now the subject of significant research interest  \cite{Optica,Davoyan_2,Shen_2,Marvasti}. Despite the huge potential of such nonreciprocity-induced hot-spots for boosting weak nonlinear effects, all studies on this topic so far have been focused on the linear response of terminated one-way channels, whereas, to the best of our knowledge, no attention has been devoted to their interactions with material nonlinearities. In the following, we propose and discuss engineered nonreciprocal plasmonic platforms to maximally enhance the field intensity at a suitable termination, while fully taking into account, for the first time, the unavoidable impact of dissipation and nonlocality. We then show that this effect can indeed lead to an improvement of several orders of magnitude in nonlinear light-matter interactions, exemplified by a giant enhancement in the efficiency of third-harmonic generation. These findings may open new directions in nonlinear electromagnetics and photonics.

\section{Giant field enhancement in terminated one-way channels}

\begin{figure*}[htb!]
	\begin{center}
		\noindent \includegraphics[width=0.62\textwidth]{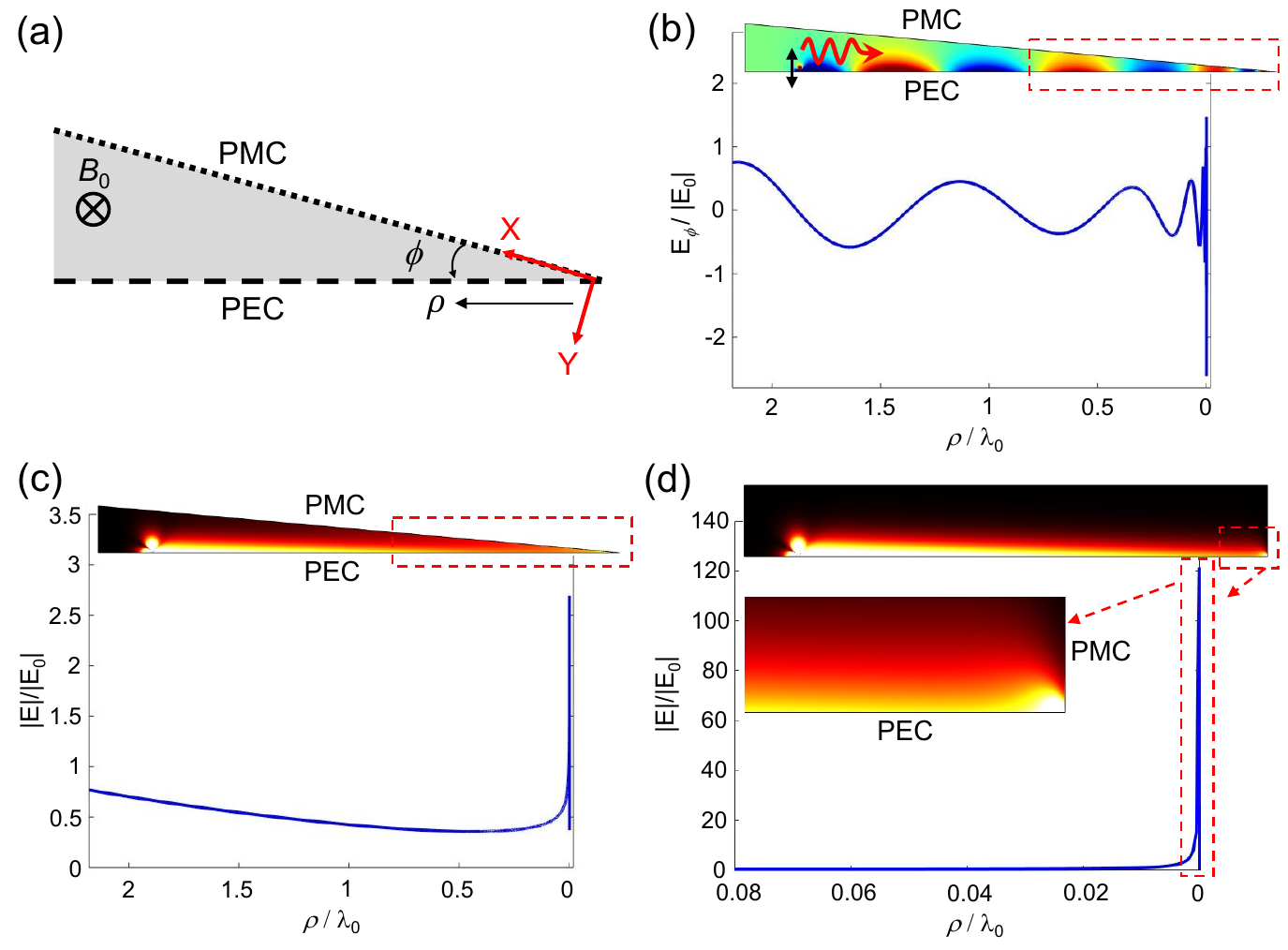}
	\end{center}
	\caption{Giant field enhancement in an idealized terminated one-way channel. (a) The geometry under consideration consists of a magnetized plasmonic wedge of angle $\phi_0 $ between perfect-electric-conducting (PEC) and perfect-magnetic-conducting (PMC) walls. (b) Time-snapshot of the normal component of the electric field at the plasma-PEC boundary, as a function of distance $\rho$ from the corner (corresponding to the dashed box in the inset). Inset: Zoomed-in view of the electric field distribution in the plasmonic wedge, launched by a point source (black arrow). The wedge angle is $ \phi_0 = 5 $ degrees, the operational frequency is $ \omega / \omega_p = 1.05 $, and the magnetized plasma is $n$-type InSb with parameters given in the text. The magnetic bias is $ B_0= 0.4  $ T, and the collision frequency is $ \Gamma/\omega_p =  0.01 $. (c) Corresponding distribution of the electric field magnitude. (d) Same as panel (c), but for a right-angle wedge with $ \phi_0 = 90 $ degrees. All the field values are normalized to $ |E_0| $, the magnitude of the electric field of a surface wave propagating along an unbounded, lossless, plasma-PEC interface.}
	\label{geom}
\end{figure*}

To gain more physical insight into the response of terminated one-way channels, we first consider an idealized configuration that is amenable to theoretical analysis. As shown in Fig. \ref{geom}(a), the structure under consideration consists of a nonreciprocal (magnetized) plasma bounded by dual hard boundaries, i.e., a perfect electric conductor (PEC) and a perfect magnetic conductor (PMC). The plasma region is biased normal to the plane, along the $z$-axis, as indicated in the figure. As discussed in \cite{Seshadri,Ishimaru,Hassani_1,Davoyan,Optica}, if the operational frequency lies within the upper bulk-mode bandgap, the interface between magnetized plasma and PEC (or a good conductor) supports a unidirectional and topological SPP mode propagating toward right or left, depending on the bias direction. This one-way propagation channel is terminated by a PMC boundary, such that a corner of angle $\phi_0$ is formed between the PEC and PMC walls, as illustrated in Fig. \ref{geom}(a). No surface mode is supported on the PMC-plasma interface, since the PMC boundary ``shorts'' the tangential magnetic field of the transverse-magnetic (TM) SPP mode, as recognized in \cite{Ishimaru} (this is strictly true only in the local case; if plasma nonlocalities are considered, an extremely confined surface mode does exist on this interface, but is very rapidly attenuated by any physical level of dissipation, as discussed in \cite{violation}). The supported one-way surface mode, therefore, cannot ``escape'' the termination since all other propagation channels -- backward propagation, radiation into the bulk, and surface-wave propagation on the PMC interface -- are forbidden. As a result, the energy carried by this mode can only accumulate at the corner, leading to a dramatic field enhancement. Indeed, the only escape channel is provided by absorption losses, which ultimately dissipate all the incident energy even in the limit of vanishing loss \cite{Ishimaru}. If the loss rate is not too large, the field intensity is expected to exhibit a sharp peak (a field hot-spot) near the corner.   

To better understand this field enhancement mechanism, we theoretically analyze the behavior of the surface mode as it approaches the corner. This analysis concerns TM modes, i.e., with $E_z = 0$, and time-harmonic dependence $ e^{-i\omega t} $. The magnetized plasmonic medium can be modeled by a non-symmetric permittivity tensor $ \boldsymbol{\epsilon} = \epsilon_0 \left[ \epsilon_{11} \textbf{I}_t + \epsilon_{33} \hat{\textbf{z}} \hat{\textbf{z}} - i\epsilon_{12}\hat{\textbf{z}} \times \textbf{I}    \right]  $, where the $z$-axis (bias direction) is supposed to be normal to the plane of propagation, $ \textbf{I}_t = \textbf{I} - \hat{\textbf{z}}\hat{\textbf{z}} $, and $ \epsilon_{12} $ is the magnitude of the gyration pseudovector. The frequency dispersion function of $ \epsilon_{11} $, $ \epsilon_{12} $ and $ \epsilon_{33} $ are \cite{Plasma}: 
\begin{align}
	& {\varepsilon _{11}} = 1 - \frac{{\omega _p^2\left( {1 + i\Gamma /\omega } \right)}}{{{{\left( {\omega  + i\Gamma } \right)}^2} - \omega _c^2}}, ~ {\varepsilon _{33}} = 1 - \frac{{\omega _p^2}}{{\omega \left( {\omega + i\Gamma } \right)}} \nonumber \\&
	{\varepsilon _{12}} = \frac{1}{\omega}\frac{{\omega _c^{}\omega _p^2}}{{\omega _c^2 - {{\left( {\omega  + i\Gamma } \right)}^2}}},
	\end{align}
	where $\omega _{p}$ is the plasma frequency, $\Gamma$ is the collision rate associated with damping, $\omega _{c}=-e|B_{0}|/m$ is the cyclotron frequency, $e$ the electron charge, $m$ the
	effective electron mass, and $B_{0}$ the static magnetic bias.

As a relevant example of solid-state magnetized plasma, we consider a magnetized semiconductor in the low THz range, e.g., $n$-type InSb with plasma frequency $ \omega_p = 2~ \mathrm{THz} $, electron density $ N_e = 1.1 \times 10^{22}/\mathrm{m^3} $, and dielectric constant due to bound charges $ \epsilon_{\infty} = 15.6 $. In this work, we suppose $|\omega_c/\omega_{p}| = 0.4$, which corresponds to a practically feasible value of bias, $ B_0 = 0.4 $ T, for the considered InSb sample. By expanding Maxwell's equations, $  \nabla \times \textbf{E} = i\omega \mu_0 \textbf{H}, ~ \nabla \times \textbf{H} = -i\omega \epsilon_0 \boldsymbol{\epsilon} \cdot \textbf{E} $ in cylindrical coordinates, it can be shown that the most general solution of the magnetic field $H_z$ takes the following form,
\begin{equation}
	H_z =\left( c_nJ_n(k_s \rho) + d_nY_n(k_s \rho)\right) \left(  a_n \cos(n\phi) + b_n \sin(n\phi)  \right),
\end{equation}
where $a_n,~b_n, ~ c_n, ~ d_n$ are unknown modal constants, and $J_n$ and $ Y_n $ are solutions of the Bessel equation of order $n$. The geometry contains the termination point $ \rho = 0 $, and the PMC and PEC boundaries are located at $ \phi = 0 $ and $ \phi = \phi_0  $, respectively. 
%
%These boundary conditions simplify the TM-mode field expressions as follows,
%\begin{align}\label{exact}
%& H_z = c_n J_n(k_s \rho) \sin(n\phi) \nonumber \\&
%E_{\rho} = \frac{c_n}{-i\omega \epsilon_0 \epsilon_{eff}} \left[  \frac{n J_n(k_s \rho)}{ k_s \rho } \cos(n\phi) + i \frac{\epsilon_{12}}{ \epsilon_{11} } J^{'}_n(k_s\rho) \sin(n \phi) \right] \nonumber \\ &
%E_{\phi} = \frac{c_n}{-i\omega \epsilon_0 \epsilon_{eff}} \left[  i \frac{\epsilon_{12}}{ \epsilon_{11} } \frac{n J_n(k_s \rho)}{ k_s \rho } \cos(n\phi) -  J^{'}_n(k_s\rho) \sin(n \phi) \right]
%\end{align}
%where $ \epsilon_{eff} = (\epsilon_{11}^2 - \epsilon_{12}^2)/\epsilon_{11} $ and $k_s = \sqrt{\epsilon_{eff}} \omega /c$ with $c$ being the vacuum speed of light. 
%
The dispersion equation of the supported surface mode can be found by applying these boundary conditions (see also Appendix \ref{APP_A}), which gives,
\begin{equation}\label{B.C.}
	\frac{n J_n(k_s \rho)}{ k_s \rho } + i \frac{\epsilon_{12}}{ \epsilon_{11} } J^{'}_n(k_s \rho) \tan(n \phi_0) = 0
\end{equation}
where $k_s = \sqrt{\epsilon_{eff}} \omega /c$ and $ \epsilon_{eff} = (\epsilon_{11}^2 - \epsilon_{12}^2)/\epsilon_{11} $, with $c$ being the speed of light in vacuum.

The allowed modal index $n$ that satisfies the dispersion equation depends on frequency and on the distance, $\rho$, from the corner. In other words, the surface mode transforms as it approaches the corner, and its amplitude, wavelength, and wave impedance depend on the allowed modal index $n$ at each distance $\rho$. Although Eq. \eqref{B.C.} is a complex, nonlinear equation and should be solved numerically, it can be simplified in the region very close to the the termination, $\rho \rightarrow 0$, by replacing the Bessel function with its small argument approximation, $ J_n(x) \simeq \frac{1}{n!}\left( \frac{x}{2} \right)^n $. This converts Eq. \eqref{B.C.} to $\tan(n \phi_0)  - i \epsilon_{11}/ \epsilon_{12} = 0$, which matches the result reported in \cite{Ishimaru}. The new dispersion equation is independent of $\rho$ since it is valid only in the extreme vicinity of the corner, and it can be solved as $n= i \tanh^{-1}(\epsilon_{11}/ \epsilon_{12})/\phi_0$, which reveals that $n$ is purely imaginary in the lossless case (since $\epsilon_{11}$ and $\epsilon_{12}$ are real).  The above assumption also simplifies the expression of the SPP electric field as follows,
\begin{align}\label{E_phi}
	E_{\phi} \simeq \frac{c_n}{-i\omega \epsilon_0 \epsilon_{eff}} n \rho^{n-1} \left[  i \frac{\epsilon_{12}}{ \epsilon_{11}} \cos(n\phi) -   \sin(n \phi) \right]
\end{align}

In a lossy structure, the modal index becomes a complex number, $n = n_r + in_i$, with $n_r >0$ in a passive system. In this case, Eq. \eqref{E_phi} shows that, as the surface wave approaches the corner, it oscillates as $ \rho^{in_i} $ and its amplitude increases/decreases as $ 1/\rho^{1-n_r} $, depending on the value of $n_r$, which in turn depends on the level of loss. Specifically, if $ 0 \leq n_r <1 $ the amplitude diverges at $ \rho = 0 $, whereas if $ n_r > 1 $ it decays to zero. Thus, this simplified analysis allows making qualitative predictions about the underdamped or overdamped behavior of surface waves in a terminated one-way channel with dissipation; however, the approximated dispersion equation and the associated field expressions do not fully capture the correct physics of a surface mode in this configuration. Specifically, the divergent behavior at $ \rho = 0 $ for a lossy structure is nonphysical since it would correspond to infinite absorbed energy. Instead, we expect the presence of a peak in the field amplitude near the termination, whose maximum value and location should depend on the level of loss and the geometry.

The correct surface-mode behavior can be captured using the exact field expressions in Eqs. \eqref{exactH}-\eqref{exactE} with the modal index $n$ numerically calculated from the exact dispersion equation, Eq. \eqref{B.C.}. The exact electric field distribution (time snapshot) for the one-way surface mode propagating toward the corner is shown in Fig. \ref{geom}(b), for the case of a lossy magnetized plasmonic taper bounded by hard boundaries with $\phi_0 = 5$ degrees. Far from the corner, the field amplitude slowly decreases due to material absorption, but, as the surface mode approaches the tip, the wavelength shrinks and the amplitude rapidly increases. Figure \ref{geom}(c) shows the magnitude of the electric field, which exhibits an evident peak near the termination. The field distributions in the figure insets clearly show that, although the mode is stopped by the termination, no backward mode is excited on either interface. 

The field behavior in this nonreciprocal plasmonic taper -- wavelength shrinking and field enhancement -- is not dissimilar from the case of reciprocal adiabatic tapers, as mentioned in the Introduction. However, there is a major difference between the two cases. In a reciprocal plasmonic taper, the surface mode energy accumulates near the tip, with minimized back reflections, only if impedance matching at any distance from the tip is ensured, which is possible for very long adiabatic tapers. Instead, in the proposed nonreciprocal plasmonic taper, impedance matching is \emph{automatically} ensured due to the inherent absence of a backward mode. This implies that adiabatic tapering is not necessary at all, and the same (or higher) level of field confinement and enhancement can be obtained for arbitrarily abrupt terminations, even for a 90 degrees corner. The field distribution for this extreme scenario is shown in Fig. \ref{geom}(d). Again, we observe that the field intensity increases dramatically close to the termination -- to even higher values than in the tapered case -- forming a clear field hot-spot as shown in the inset. We also note that, despite the abrupt, non-adiabatic termination, such a field enhancement effect occurs over the entire frequency window in which unidirectional surface wave propagation is supported, which can be quite wide depending on the magnetic bias intensity \cite{Optica,Fan,TAP}.

%Although the structure supports unidirectional mode at the fundamental frequency, the system seen by the higher harmonic has different properties than the fundamental frequency due to the dispersive nature of the material. In this situation, by properly tuning the frequencies, the higher harmonic signal can be filtered out from the fundamental frequency signal and guided backward. 

\section{Broadband and Enhanced Nonlinear Effects}

The broadband giant hot-spots supported by terminated one-way channels appear ideal to enhance weak light-matter interactions, especially nonlinear effects since they depend on the powers of the local field. However, the idealized configuration in Fig. \ref{geom}, which uses PEC and PMC boundaries, is not practical, especially at frequencies above the microwave range. %The need to hard boundaries comes from the fact that the unidirectional surface wave within the bulk mode band gap frequency range (in-the-gap surface wave) emerges on the interface with an opaque medium. 
Fortunately, as mentioned in the Introduction, unidirectional surface waves known as surface magneto-plasmons also exist on an interface between a magnetized plasma and a dielectric material, for frequencies below the plasma frequency \cite{1972,Optica,diffractionless,Fan,TAP}. An important advantage of these surface waves is that there is no need to impose impractical PMC boundaries to create a termination. In fact, surface magneto-plasmons can be stopped by a PEC wall or by an interface with a conventional opaque medium, as discussed in \cite{Optica,Fan,Davoyan_2}. A disadvantage of this configuration is that a backward mode may be excited if nonlocal effects are included in the material model, and the impact of this additional propagation channel must be assessed carefully, as discussed below.

\begin{figure}[h!]
	\begin{center}
		\noindent \includegraphics[width=0.85\columnwidth]{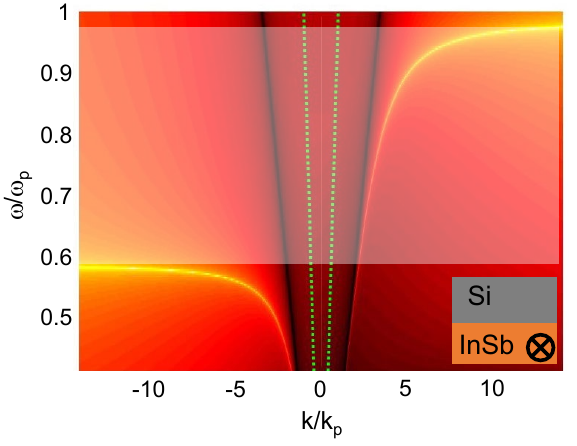}
	\end{center}
	\caption{Dispersion diagram of the surface modes supported by the configurations in the inset. The dispersion diagram is plotted as a density plot of the inverse determinant of the boundary-condition matrix. The bright bands correspond to the SPP poles. The magnetized plasma is InSb with parameters given in the text and cyclotron frequency $ |\omega_c|/\omega_p = 0.4 $. The green dashed lines represent the light lines for plane waves propagating in free space. The shaded white area indicates the frequency window where unidirectional surface modes are supported.}
	\label{nonlinear}
\end{figure}

We first consider a structure consisting of semi-infinite layers of biased InSb and silicon (Si), with relative permittivity $ \epsilon_d = 11.68 $ \cite{Randall}. The SPP dispersion diagram for this structure is shown in Fig. \ref{nonlinear}. The unidirectional frequency window of surface magneto-plasmons, indicated by the shaded white area in the figure, is defined by the following upper and lower bounds: $ \omega_{\pm} = ( \pm \omega_c + \sqrt{ 2\omega_p^2 + \omega_c^2 } )/2 $ \cite{diffractionless,Fluctuation}. When $ \omega_c = 0 $, the unidirectional frequency window closes and the interface supports symmetric and bidirectional SPPs.

\begin{figure*}[hbt!]
	\begin{center}
		\noindent \includegraphics[width=0.7\textwidth]{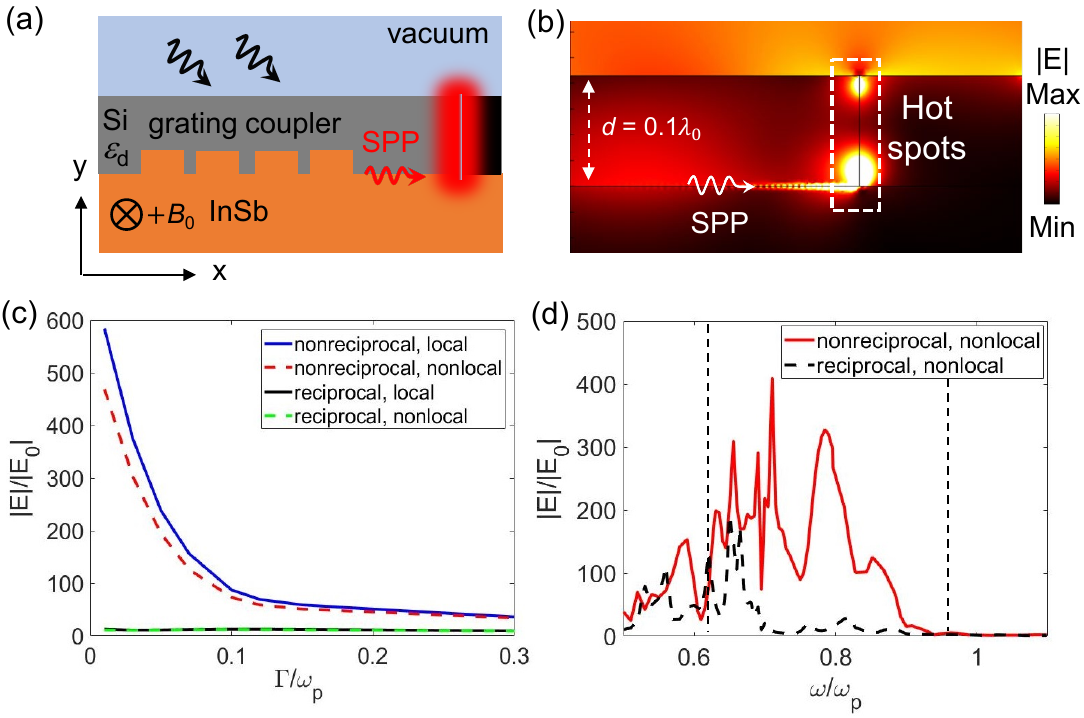}
	\end{center}
	\caption{Broadband field enhancement, and impact of dissipation and nonlocality. (a) Illustration of the terminated one-way waveguiding geometry under consideration. A one-dimensional grating is introduced to couple the incident propagating wave to the one-way surface mode. The grating is composed of six ridges of width 32 $\mu $m $= 0.15\lambda_0 $, height 10.7 $ \mu $m $= \lambda_0/20 $ and period 64 $\mu $m $= 0.3\lambda_0 $, where $ \lambda_0 $ is the free-space wavelength at $ \omega/\omega_p = 0.71 $. The termination region (black) is composed of an isotropic metal with $ \text{Re}[\epsilon_m] = -\epsilon_d $, at $ \omega / \omega_p = 0.71 $. (b) Distribution of the electric field magnitude for the geometry in panel (a), with the same InSb parameters as in Fig. \ref{geom}. The Si layer has thickness $ d = 0.1 \lambda_0 $. (c) Field enhancement at the termination for different scenarios (reciprocal/nonreciprocal and local/nonlocal) as a function of the damping rate, $\Gamma$, at a fixed frequency, $ \omega / \omega_p = 0.71 $. For the nonlocal cases, $ \beta = 1.07 \times 10^{6} $ m/s \cite{Fan}. (d) Field enhancement as a function of frequency, for a fixed level of loss, $ \Gamma/\omega_p = 0.01 $. Vertical dashed lines indicate the unidirectional frequency window, consistent with Fig. \ref{nonlinear}. The field is normalized at each frequency with respect to the electric field $|E_0|$ of a surface wave propagating along a low loss ($  \Gamma/\omega_p = 0.01 $), local, non-terminated biased structure. For all panels, the structure is illuminated by a Gaussian beam with waist $ w_0 = 100 ~ \mu \mathrm{m} = 2.14 \lambda_0 $ and incident angle of $ \theta = 45 $ degrees with respect to the interface. While these simulations were performed in 2D, a representative 3D simulation is available in the Supplemental Material \cite{SM}, demonstrating the excitation of unidirectional SPPs by a 3D Gaussian beam and the emergence of an extended electric field hot-spot.}
	\label{nonlinear2}
\end{figure*}

We note that, as seen in Fig. \ref{nonlinear}, the unidirectional SPP dispersion curves lie below the light cone for plane waves propagating in free-space (green dashed lines), which implies that the mode requires an evanescent-wave excitation (e.g., a near-field point source), as in the case of conventional SPPs. A more practical configuration that allows the one-way surface mode to be excited by an incident propagating wave, e.g., a laser beam, can be obtained using well-established methods, for example realizing a grating coupler by periodically corrugating the surface of the plasmonic medium. As usually done, the period $P$ is chosen such that the grating compensates the momentum mismatch between the surface mode and a free-space propagating wave (one of the space harmonics of the SPP mode ends up within the light cone): the new SPP wavevector becomes $k'=k+2 \pi n/P$, where $2 \pi n/P$ is a reciprocal lattice vector \cite{grating}. Then, if the silicon layer is interfaced with free space, as shown in Fig. \ref{nonlinear2}(a), the unidirectional surface magneto-plasmon becomes, within the grating coupler region, a one-way ``leaky'' mode that can couple to, and be excited by, free-space propagating waves \cite{FM_leaky,JPCM}. Additional details about the geometry are provided in the caption of Fig. \ref{nonlinear2}. An alternative configuration that does not require a grating coupler is presented in \cite{SM}.

The opaque termination that is introduced to stop the excited surface magneto-plasmons is indicated by the black region in Fig. \ref{nonlinear2}(a). To boost the field enhancement further, a resonant termination may be designed, for example in the form of a non-magnetized plasmonic material with a Drude-like permittivity such that $ \text{Re}[\epsilon_m] = -\epsilon_d = -11.6  $ at the central frequency of interest. Fig. \ref{nonlinear2}(b) shows the electric field magnitude distribution in this structure, under external plane-wave illumination. One-way surface waves are launched by the grating toward right and giant field hot-spots are clearly visible at the termination.

As mentioned at the beginning of this section, a disadvantage of this configuration is its fragility to nonlocal effects (spatial dispersion), which is a fundamental feature of real plasmonic materials. Indeed, as comprehensively discussed in \cite{Optica, Fan}, surface magneto-plasmons are not strictly unidirectional if nonlocality is included in the material model. In the nonlocal scenario, a backward-propagating surface mode emerges within the unidirectional frequency window. While this makes the system no longer strictly unidirectional, the dispersion diagram remains strongly asymmetrical, as the backward-propagating mode exists for large wavenumber values and is rapidly attenuated for moderate levels of dissipation. Thus, the relative impact of dissipation and nonlocality should be carefully assessed in order to make correct predictions regarding the maximum field intensity at the termination. Nonlocal effects can be included in the InSb material model by writing Ampere's Law as $ \nabla \times \textbf{H} = -i\omega \epsilon_0 \epsilon_{\infty} \textbf{E} + \textbf{J} $, where $ \textbf{J} $ is the induced free-electron current governed by a hydrodynamic equation of motion \cite{Optica,Fan,Raza,Smith_Microscopic}, 
\begin{equation}\label{hydro}
\beta^2 \nabla ( \nabla \cdot \textbf{J} ) + \omega ( \omega + i\Gamma ) \textbf{J} = i\omega \left(  \omega_p^{2} \varepsilon_0 \varepsilon_{\infty} \textbf{E}  - \textbf{J} \times \omega_c \hat{z} \right)
\end{equation}
where $ \beta = 1.07 \times 10^{6} $ m/s is the nonlocal parameter and $ \Gamma $ is the damping rate due to absorption losses. The first term of the equation is a pressure term determining the convective currents that are responsible for nonlocal effects. The silicon layer is instead assumed to be local, as usually done for dielectric materials \cite{Raza}. The field distribution everywhere can then be calculated numerically using the finite element method \cite{comsol}, solving Maxwell's equation and the hydrodynamic equation simultaneously, with suitable boundary conditions. In particular, neglecting electron spill-over, the free-electron current normal to the surface is required to vanish at the plasma-dielectric interface \cite{Optica,Fan,Raza}. 

To understand how the presence of dissipation and nonlocality affects the field hot-spots, we investigate the field enhancement at the termination for different scenarios. Specifically, Fig. \ref{nonlinear2}(c) compares the field enhancement for reciprocal/nonreciprocal and local/nonlocal InSb as a function of the damping rate, $\Gamma$, in the magnetized plasmonic material, at a frequency within the unidirectional frequency window, $ \omega / \omega_p = 0.71 $. These results confirm that the field enhancement achievable in the nonreciprocal structure is an order of magnitude larger than in the reciprocal case ($\omega_c=0$), even for large values of dissipation. In the reciprocal configuration, a simple standing wave forms along the channel, %formed by counter-propagating SPPs with identical (opposite) wavenumbers, 
which only produces a moderate field enhancement. Importantly, we see that nonlocal effects do reduce the enhancement due to the emergence of a new backward-propagation channel \cite{Optica,Fan}; however, as mentioned above, the surface mode dispersion remains strongly asymmetrical and only SPPs with very large wavevectors can escape through this channel. As a result, despite the presence of spatial dispersion, the field enhancement remains an order of magnitude larger than in the reciprocal case, even in the presence of substantial optical losses. In Fig. \ref{nonlinear2}(d), we also show this enhancement over a wide range of frequencies, for both the reciprocal and nonreciprocal nonlocal cases. These results clearly demonstrate that, in sharp contrast to the conventional enhancement methods based on resonances, a very large electric field intensity is achievable here over the entire unidirectional frequency window. The oscillations in Fig. \ref{nonlinear2}(d) are mostly due to the dispersive nature of the involved materials. The physical reason for this broadband effect is that no resonance is required in the proposed process (although a resonant termination is employed to further increase the field enhancement). This behavior is ideal to boost light-matter interactions and nonlinear effects in a broadband fashion. In the following, we theoretically demonstrate the potential of these ideas, for a specific nonlinear process, by investigating the third-harmonic generation (THG) efficiency in the proposed waveguiding platform, considering the natural $\chi^{(3)}$ nonlinear properties of silicon.

In materials with non-negligible third-order nonlinear susceptibility, a third-harmonic (TH) wave is generated by the nonlinear polarization density $ \textbf{P}_{TH} = \epsilon_0 \chi^{(3)} \textbf{E}_{FF}^3 $, where $ \textbf{E}_{FF}$ is the electric field phasor at the fundamental frequency, equal to the frequency of the Gaussian beam illuminating the structure. %The nonlinear polarization is then included as a new source term in the wave equation describing the nonlinear material regions: $ \nabla \times \nabla \times \textbf{E} - \frac{n^2}{c^2} \frac{\partial^2 \textbf{E} }{\partial t^2}= - \frac{1}{\epsilon_0 c^2} \frac{\partial^2  \textbf{P}_{TH}}{ \partial t ^2  } $. 
Silicon is an indirect bandgap semiconductor with a pronounced third-order nonlinear susceptibility on the order of $ \chi^{(3)} = 2.8 \times 10^{-18} ~ \mathrm{m^2 / V^2} $, while for the metallic termination we consider the third-order nonlinear susceptibility of silver: $ \chi^{(3)} = 3 \times 10^{-19} ~ \mathrm{m^2 / V^2} $ \cite{Boyd}. Instead, the nonlinear susceptibility of InSb is much lower and is neglected. As discussed in the Supplemental Material \cite{SM}, under the undepleted pump approximation, we can use the linear frequency-domain wave equation to solve for the steady state fields at the third-harmonic frequency. This is done by simply including a new source term proportional to the nonlinear polarization density, $\mu_0 \omega^2 \textbf{P}_{TH}$, and then solving the wave equation at the third-harmonic frequency using the finite element method \cite{comsol}. This modeling approach is consistent with recent works on nonlinear metamaterials and nanophotonics \cite{SM_1,SM_2,SM_3,SM_4}, which have shown this method to produce results in good agreement with experiments. In addition to these nonlinear effects, we still fully include nonlocalities and absorption losses in the InSb region, making the system under study nonlinear, nonlocal, nonreciprocal, dissipative, and dispersive.

The strength of the THG process is evaluated by computing the conversion efficiency defined as $ \mathrm{CE} = \mathrm{P}_{out,TH}/ \mathrm{P}_{in, FF} $, that is, as the ratio between the output power at the third-harmonic frequency and the input power at the fundamental frequency. The input power of the incident Gaussian beam can be calculated as $ \mathrm{P}_{in, FF} = 0.5 \pi I_0 w_0^2 \cos ( \theta_0 ) $, where $ I_0 = H_0^2 \eta/2 $ is the maximum beam intensity with $ H_0 $ being the magnitude of the incident magnetic field and $ \eta = 377 ~ \Omega $ the free space impedance. The total output power is computed by integrating the outgoing Poynting vector on the outer boundaries of the entire computational domain at the third-harmonic frequency. While the structure supports guided unidirectional modes at the fundamental frequency, the system operating at the third harmonic has different properties due to the dispersive nature of the materials involved. In particular, since the third-harmonic frequency is much larger than the plasma frequency of the nonreciprocal plasmonic material, this medium is mostly transparent to the third-harmonic field, which is therefore able to exit the structure, while the fundamental-frequency field is trapped in the waveguiding structure. % In this situation, by properly tuning the frequencies, the higher harmonic signal can be filtered out from the fundamental frequency signal and guided backward. 

\begin{figure*}[hbt!]
	\begin{center}
		\noindent \includegraphics[width=0.7\textwidth]{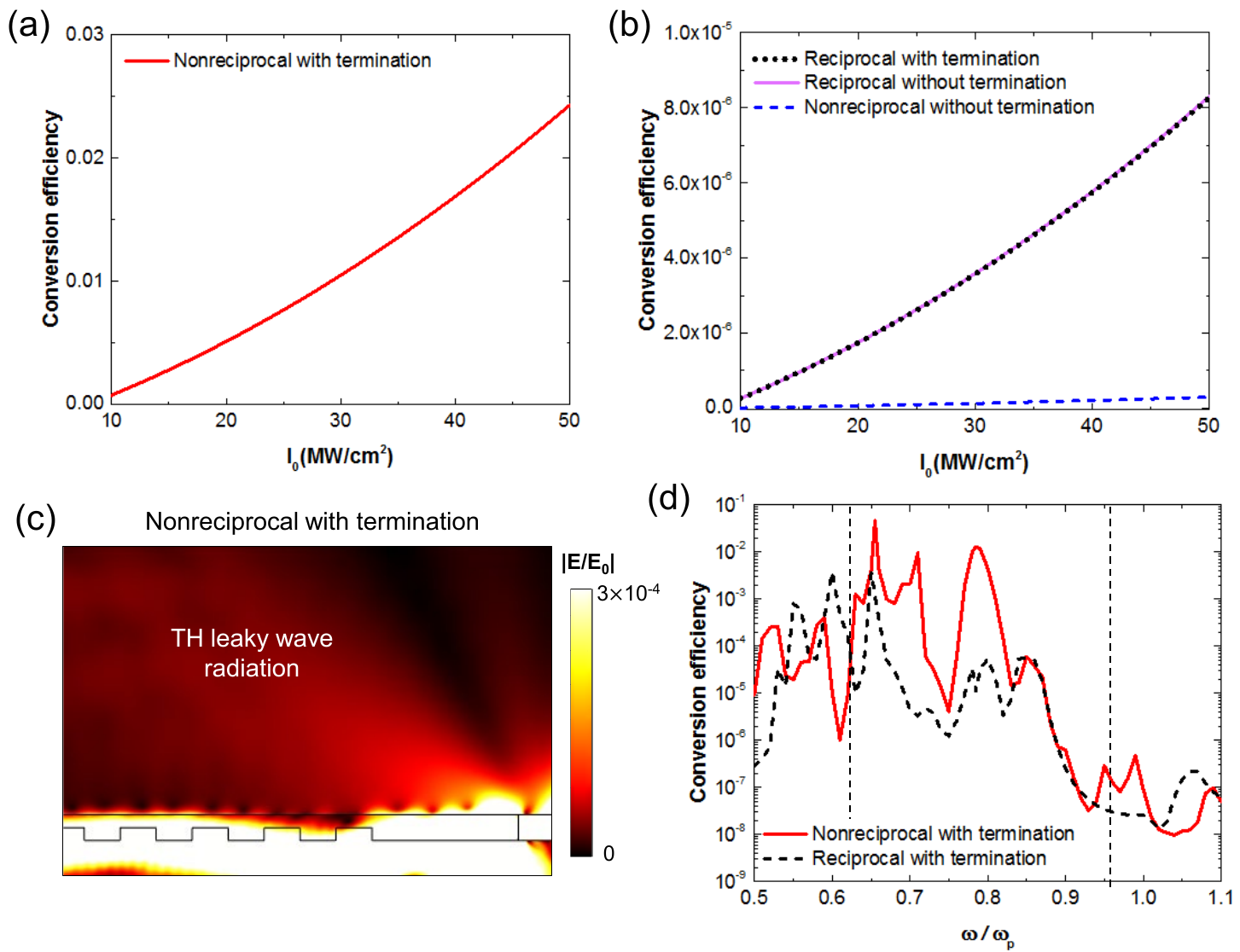}
	\end{center}
	\caption{Third-harmonic-generation (THG) conversion efficiency for (a) the nonreciprocal plasmonic waveguide with metallic termination considered in Fig. \ref{nonlinear2} (red solid line); and (b) the same nonreciprocal waveguide, but without termination (blue dashed line), and the same waveguide, but without bias, with termination (black dotted line) and without termination (purple solid line). (c) Distribution of the electric field magnitude at the third-harmonic frequency for the terminated nonreciprocal waveguide. Due to the dispersive nature of the materials, the waveguide becomes only weakly nonreciprocal at the third-harmonic frequency and the THG field can escape the structure in the form of leaky-wave radiation. In this panel, the incident wave peak intensity is $50~ \mathrm{MW/cm^2}$, low enough to avoid any heat damage to the involved materials. Panels (a)-(c) are calculated for a fundamental frequency of $ \omega/\omega_p = 0.71 $. (d) THG conversion efficiency as a function of fundamental frequency, for a fixed input intensity $I_0 = 50$ MW/cm$^2$. As in Fig. \ref{nonlinear2}(d), vertical dashed lines indicate the unidirectional frequency window, consistent with Fig. \ref{nonlinear}.}
	\label{fig3}
\end{figure*}

The computed THG $\mathrm{CE}$ results are reported in Fig. \ref{fig3}(a) for the terminated one-way waveguiding structure in Fig. \ref{nonlinear2}. Rather remarkably, the THG conversion efficiency takes very high values, on the order of few percent, by using relatively low input intensities. This is indeed due to the giant field enhancement and confinement of the fundamental-frequency wave at the termination. In comparison, the $ \mathrm{CE} $ is more than \emph{four} orders of magnitude lower in the nonreciprocal case without termination, and in the reciprocal cases with or without termination, as shown in Fig. \ref{fig3}(b). We also note that input intensity values comparable to the ones considered here have been experimentally obtained at infrared frequencies by using different excitation configurations emitting terahertz pulses, as well as using quantum cascade lasers \cite{Jepsen,You,Zhao,Natale}. The relatively low input intensity values considered here also ensure that we operate within the undepleted pump regime \cite{Boyd}, and that detrimental thermal effects will not affect the waveguide performance. Specifically, we show in \cite{SM} that the materials of the proposed structure can withstand the absorption-induced heating for input intensities lower than $I_0 = 60$ MW/cm$^2$, as in Fig. \ref{fig3}. In addition, even if much lower input intensities were considered, our structure would still exhibit a huge enhancement in conversion efficiency compared to the reciprocal case. Finally, we note that much larger TH fields can be generated, with similar efficiencies, using arrays of terminated one-way channels forming large-scale interfaces, as shown in the Supplementary Material \cite{SM}.

The computed field distribution at the third-harmonic frequency is shown in Fig. \ref{fig3}(c) for the nonreciprocal terminated waveguide. At this frequency, the structure becomes only weakly nonreciprocal, and the TH wave is therefore allowed to propagate backward from the termination and escape the structure in the form of radiation from a well-defined leaky wave (while the output TH power leaks mainly downward, if desired it can be directed upward by simply introducing a suitable mirror right below the structure). The TH leaky-wave radiation is also moderately directive, and could be made even more directive with a suitable structure, as in conventional leaky-wave antennas \cite{FM_leaky}. The field distributions for all the different configurations considered in Fig. \ref{fig3}(b) -- reciprocal/nonreciprocal and with/without termination -- are provided in Supplemental Material \cite{SM}. Most importantly, Fig. \ref{fig3}(d) shows the THG conversion efficiency over a wide frequency range, for the terminated reciprocal and nonreciprocal waveguides, demonstrating a broadband orders-of-magnitude improvement in the nonreciprocal case. A comparison with Fig. \ref{nonlinear2}(d) shows that such a broadband nonlinear effect indeed originates from the broadband linear field enhancement in our structure. These results clearly demonstrate the potential of the proposed terminated one-way plasmonic waveguides to achieve extreme levels of field enhancement and giant nonlinear effects over a broad range of frequencies.

%\red{[if you agree, I'm thinking to remove this paragraph to streamline the text, and just refer to the supplementary material]Without the metallic termination, the third harmonic field is strongest just below the incident point, at the center of the silicon layer, and radiates out into the free space from both up and down directions, as it can be seen in the supplementary material where we plot the third harmonic field distribution for all reciprocal and nonreciprocal structure with and without termination cases. With the metallic termination, the fundamental field is dramatically enhanced on the silicon-metal boundary only in the biased configuration. This field enhancement, along with an increase in the intensity of the input beam, triggers strongly enhanced nonlinear optical effects leading to the presented boosted third harmonic generation. } 

%\red{[Boyuan, Chris, I guess that the different panels in Figs. S1 and S3 have different color range, right? Let's have color bars for all panels. Let's also remove the additional lines in the plots near the termination (the artificial lines due to local meshing near the termination)].}

\section{Conclusion}

In summary, in this article we have proposed and theoretically demonstrated a new strategy to realize remarkably strong, broadband, nonlinear light-matter interactions. This strategy is not based on resonant mechanisms (e.g., localized surface-plasmon resonances) or adiabatic impedance matching in slow-light structures (e.g., plasmonic tapers), but rather on the extreme electromagnetic response of terminated one-way nonreciprocal channels. Specifically, we have shown that the proposed structure supports giant, broadband, field hot-spots, even if losses and nonlocalities are fully taken into account, and we have numerically demonstrated that such hot-spots can be exploited to achieve an enhancement of several orders of magnitude in nonlinear processes, for example third-harmonic generation, over a small footprint.

Our results show the potential of combining strongly nonreciprocal (unidirectional) platforms with optical nonlinear effects. More broadly, the structures considered in this work, which are, at the same time, \emph{nonreciprocal, nonlocal, nonlinear and non-Hermitian} (due to absorption and radiation losses), offer a glimpse into the rich physics of exotic electromagnetic and photonic platforms, which provide new opportunities that are only now beginning to be explored. 

\section*{Acknowledgements}
F.M. and S.A.H.G. acknowledge support from the Air Force Office of Scientific Research with Grant No. FA9550-19-1-0043 and the National Science Foundation (NSF) with Grant No. 1741694. C.A. and B.J. have been partially supported by the Office of Naval Research Young Investigator Program (ONR-YIP) Award with Grant No. N00014-19-1-2384, and the NSF‐Nebraska‐EPSCoR with Grant No. OIA-1557417.

\appendix

\section{SPP modal analysis of a terminated one-way channel}
\label{APP_A}

Consider the geometry in Fig. \ref{geom}(a), with the nonreciprocal plasmonic material (magnetized $n$-type InSb) characterized by a gyrotropic permittivity tensor. Starting from Maxwell's equation in cylindrical coordinates, we obtain the following equations for the different components of the supported transverse magnetic (TM) mode at the InSb-PEC interface

\begin{align}\label{basics}
& \frac{1}{\rho} \left[  \frac{\partial(\rho E_{\phi})}{ \partial \rho}   - \frac{  \partial E_{\rho}  }{ \partial \phi } \right] = i\omega \mu_0 H_z \nonumber \\&
\frac{1}{ \rho } \frac{  \partial H_z }{ \partial \phi } = -i\omega \epsilon_0 \left[  \epsilon_{11} E_{\rho} + i \epsilon_{12} E_{\phi}  \right] \nonumber \\&
- \frac{  \partial H_z }{ \partial \rho } = -i\omega \epsilon_0 \left[ -i \epsilon_{12} E_{\rho} +  \epsilon_{11} E_{\phi}  \right].
\end{align}

Using these equations, $H_z$ can be obtained in terms of the other field components,
\begin{align}\label{bulk_1}
& \frac{1}{ \rho } \frac{  \partial^2 H_z }{ \partial \phi^2 } +  \frac{\partial }{  \partial \rho } \left[  \rho \frac{  \partial H_z }{ \partial \rho }  \right] = \nonumber \\&
 -i\omega \epsilon_0 \left[   \epsilon_{11} \left( \frac{ \partial E_{\rho} }{  \partial \phi } -   \frac{ \partial  (\rho E_{\phi})}{ \partial \rho }  \right) + i\epsilon_{12} \left(   \frac{ \partial E_{\phi}  }{ \partial \phi } +  \frac{ \partial (\rho E_{\rho}) }{  \partial \rho } \right)  \right].
\end{align}

The transversality condition (Gauss Law), $ \nabla \cdot \textbf{D} = \nabla \cdot (\boldsymbol{\epsilon} \cdot \textbf{E}) = 0 $, gives 
\begin{equation}\label{div}
\epsilon_{11} \left[ \frac{ \partial( \rho E_{\rho} ) }{ \partial \rho } + \frac{  \partial E_{\phi} }{ \partial \phi }  \right] + i\epsilon_{12} \left[  \frac{ \partial ( \rho E_{\phi} ) }{ \partial \rho }  - \frac{ \partial E_{\rho} }{ \partial \phi } \right] = 0.
\end{equation}
The first equation in \eqref{basics} can be written as 
\begin{equation}\label{transversality_1}
\left[  \frac{ \partial ( \rho E_{\phi} ) }{ \partial \rho }  - \frac{ \partial E_{\rho} }{ \partial \phi } \right] = i\omega \mu_0 \rho H_z,
\end{equation}
which, combined with Eq. \eqref{div}, gives
\begin{equation}\label{transversality_2}
\frac{  \partial (\rho E_{\rho} )  }{ \partial \rho } + \frac{ \partial E_{\phi} }{ \partial \phi } = -i\frac{\epsilon_{12}}{ \epsilon_{11} } \left[ \frac{\partial(\rho E_{\phi})}{ \partial \rho}   - \frac{  \partial E_{\rho}  }{ \partial \phi }   \right] = \frac{\epsilon_{12}}{ \epsilon_{11} } \omega \mu_0 \rho H_z.
\end{equation}
Finally, substituting Eqs. \eqref{transversality_1} and \eqref{transversality_2} in \eqref{bulk_1} gives the following differential equation for $H_z$,

\begin{equation}
\frac{1}{ \rho } \frac{  \partial^2 H_z }{ \partial \phi^2 } +  \frac{\partial }{  \partial \rho } \left[  \rho \frac{  \partial H_z }{ \partial \rho }  \right] = -k_0^2 \epsilon_{eff} \rho H_z.
\end{equation}

By inserting a solution of the form $H_z = A(\rho) B(\phi)$ in the above equation, and defining $k_s^2 = k_0^2 \epsilon_{eff} $, we obtain,
\begin{equation}
\frac{1}{B} \frac{  \partial^2 B }{ \partial \phi^2 } + \frac{1}{A}  \rho \frac{\partial}{ \partial \rho } \left[ \rho \frac{ \partial A }{ \partial \rho } \right] + k_s^2 \rho^2 = 0,
\end{equation}
which gives the following ODEs for the radial and angular terms,
\begin{align}
& \frac{1}{B} \frac{  \partial^2 B }{ \partial \phi^2 } =- n^2 \nonumber \\&
\rho \frac{\partial}{ \partial \rho } \left[ \rho \frac{ \partial A }{ \partial \rho } \right] + \left(k_s^2 \rho^2 - n^2\right)A = 0,
\end{align}
with the following general solutions
\begin{align}
& B = a_n \cos(n\phi) + b_n \sin(n\phi) \nonumber \\ &
A = c_nJ_n(k_s \rho) + d_nY_n(k_s \rho).
\end{align}
Thus, $H_z$ takes the following form,
\begin{equation}
H_z =\left( c_nJ_n(k_s \rho) + d_nY_n(k_s \rho)\right) \left(  a_n \cos(n\phi) + b_n \sin(n\phi)  \right).
\end{equation}

The other field components can be written in terms of $H_z$ as,
\begin{align}
& E_{\rho} = \frac{1}{-i\omega \epsilon_0 \epsilon_{eff}} \left[ \frac{1}{\rho} \frac{ \partial H_z }{ \partial \phi } + i\frac{\epsilon_{12}}{\epsilon_{11}} \frac{ \partial H_z }{ \partial \rho }  \right] \nonumber \\ &
E_{\phi} = \frac{1}{-i\omega \epsilon_0 \epsilon_{eff}} \left[ i\frac{\epsilon_{12}}{\epsilon_{11}} \frac{1}{\rho} \frac{ \partial H_z }{ \partial \phi } -  \frac{ \partial H_z }{ \partial \rho }  \right].
\end{align}

The geometry contains the point $ \rho = 0 $ (wedge apex), which implies that $ d_n =0 $, otherwise the field solution would diverge in all cases. In addition, the PMC boundary is at $ \phi = 0 $, which implies that $ a_n = 0 $ (vanishing tangential magnetic field on the PMC boundary). Applying these boundary conditions simplifies the field components as follows, 
\begin{equation} \label{exactH}
H_z = c_n J_n(k_s \rho) \sin(n\phi)
\end{equation}
\begin{align} \label{exactE}
& E_{\rho} = \frac{c_n}{-i\omega \epsilon_0 \epsilon_{eff}} \left[  \frac{n J_n(k_s \rho)}{ k_s \rho } \cos(n\phi) + i \frac{\epsilon_{12}}{ \epsilon_{11} } J^{'}_n(k_s\rho) \sin(n \phi) \right] \nonumber \\ &
E_{\phi} = \frac{c_n}{-i\omega \epsilon_0 \epsilon_{eff}} \left[  i \frac{\epsilon_{12}}{ \epsilon_{11} } \frac{n J_n(k_s \rho)}{ k_s \rho } \cos(n\phi) -  J^{'}_n(k_s\rho) \sin(n \phi) \right]
\end{align}

Finally, since the PEC boundary is located at $ \phi = \phi_0 $, the tangential electric field vanishes on this boundary, $ E_{\rho}\lvert_{\phi = \phi_0} = 0 $, which leads to the following dispersion equation for the TM modes supported by the wedge-like structure in Fig. \ref{geom}(a)
\begin{equation}
\frac{n J_n(k_s \rho)}{ k_s \rho } + i \frac{\epsilon_{12}}{\epsilon_{11}} \frac{\partial J_n(k_s \rho)}{ \partial (k_s \rho) } \tan(n \phi_0) = 0,
\end{equation}
which corresponds to Eq. \eqref{B.C.} of the main text.

%%%%%%%%%%%%%%%%%%%%%%%%%%%%%%%%%%%%%%%%%%%%%%
%%%%%%%%%%%%%%%%%%%%%%%%%%%%%%%%%%%%%%%%%%%%%%
%%%%%%%%%%%%%%%%%%%%%%%%%%%%%%%%%%%%%%%%%%%%%%

\end{document}